# Manifestations of the Charged Stripes in the Magnetoresistance of Heavily Underdoped $YBa_2Cu_3O_{6+x}$

A. N. Lavrov*, Yoichi Ando, and Kouji Segawa

Central Research Institute of Electric Power Industry, 2-11-1 Iwado-kita, Komae, Tokyo 201-8511, Japan

**Abstract:** We present a study of the in-plane and out-of-plane magnetoresistance (MR) in heavily-underdoped, antiferromagnetic $YBa_2Cu_3O_{6+x}$, which reveals a variety of striking features. The in-plane MR demonstrates a "*d*-wave"-like anisotropy upon rotating the magnetic field $H$ within the *ab* plane. With decreasing temperature below 20-25 K the system acquires memory: exposing a crystal to the magnetic field results in a persistent in-plane resistivity anisotropy. The overall features can be explained by assuming that the $CuO_2$ planes contain a developed array of stripes accommodating the doped holes, and that the MR is associated with the field-induced topological ordering of the stripes.

**Keywords:** Stripes, Magnetoresistance, Antiferromagnetic, YBaCuO

## INTRODUCTION

In high-$T_c$ cuprates the conducting state appears as a result of hole or electron doping of the parent antiferromagnetic (AF) insulator. The tendency of doped holes to segregate may give rise to an intriguing microscopic state with carriers gathered within an array of quasi-1D "stripes" separating AF domains [1-3]. An ordered striped structure has been observed [4] in $La_{1.6-x}Nd_{0.4}Sr_xCuO_4$ and in $La_2NiO_{4.125}$, while most superconducting cuprates demonstrate incommensurate magnetic fluctuations [5] which can be considered as dynamical stripe correlations [1]. Dynamical or static stripes might be responsible for the peculiar normal state of cuprates as well as for the occurrence of superconductivity [1], but still very little is known about the electron dynamics in the stripes.

In this paper we report an extraordinary behavior of the magnetoresistance (MR) in antiferromagnetic $YBa_2Cu_3O_{6+x}$, which provides evidences that conducting stripes actually exist in $CuO_2$ planes and have a considerable impact on the electron transport.

## EXPERIMENTAL METHODS

The high-quality $YBa_2Cu_3O_{6+x}$ single crystals were grown by the flux method in $Y_2O_3$ crucibles, and a high-temperature annealing was used to reduce their oxygen content. The MR was measured by sweeping the magnetic field at fixed temperatures stabilized by a capacitance sensor with an accuracy of ~1 mK. The angular dependence of the MR was determined by rotating the sample within a 100$^o$ range under constant magnetic fields up to 16 T.

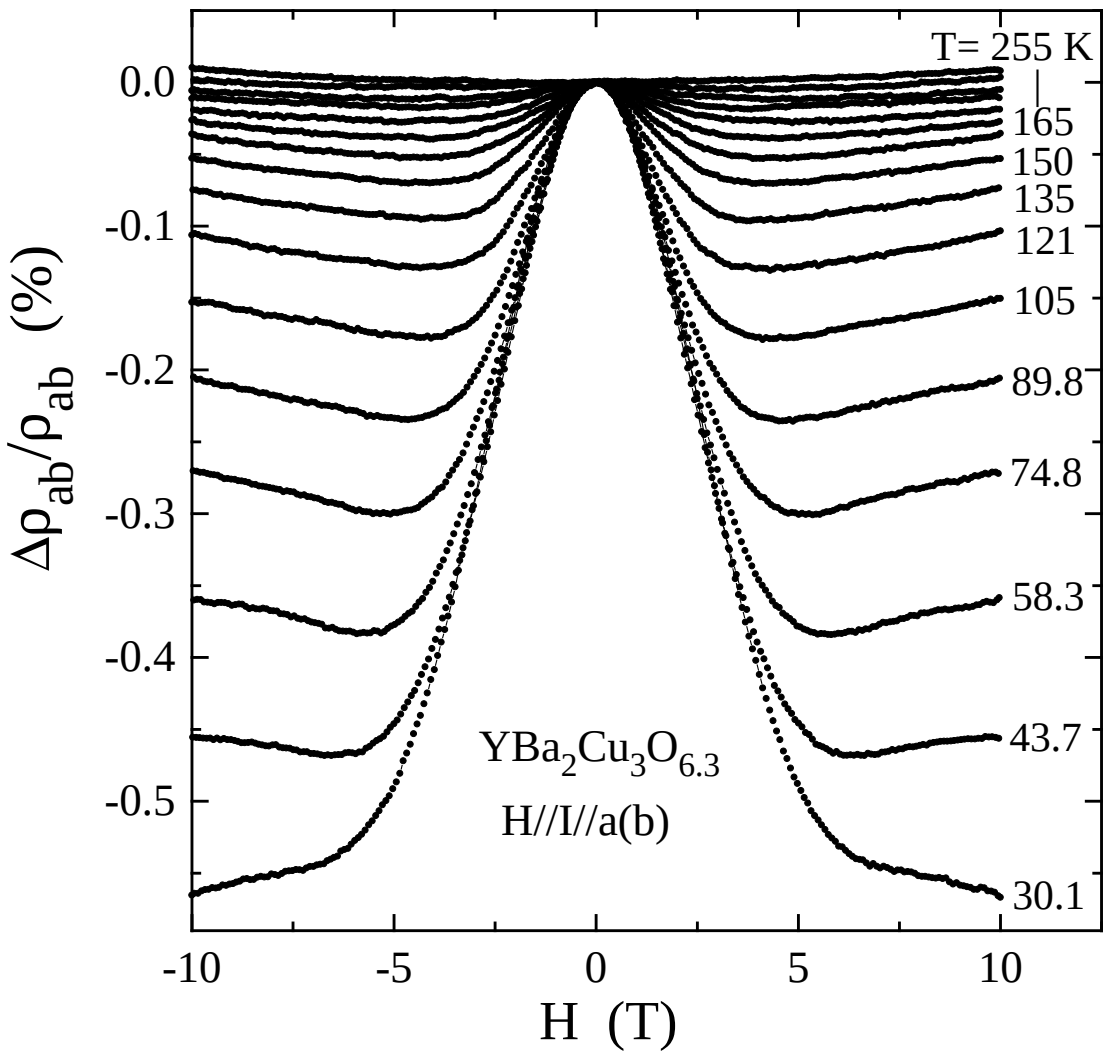

**Fig. 1.** Longitudinal in-plane MR of $YBa_2Cu_3O_{6.3}$. The data are averaged over several field sweeps.

## RESULTS AND DISCUSSION

The $YBa_2Cu_3O_{6+x}$ crystals even being located deep in the AF range of the phase diagram ($x$~ 0.3) are far from conventional insulators: the in-plane resistivity $\rho_{ab}$ remains "metallic" at high $T$ and at low $T$ it grows slower than expected for the hopping electron transport [6,7]. These AF crystals demonstrate an unusual behavior of the in-plane MR, $\Delta\rho_{ab}/\rho_{ab}$, when the magnetic field $H$ is applied along the $CuO_2$ planes, Fig.1. At weak fields, the longitudinal in-plane MR [$H//I//ab$] is negative and follows roughly a $T$-independent $\zeta H^2$ curve, but abruptly saturates above some threshold field. The threshold field and the saturated MR value gradually increase with decreasing temperature. The MR anomaly becomes noticeable near the Néel temperature $T_N \approx 230$ K, but evolves rather smoothly through $T_N$, which indicates that the long-range AF order itself is not responsible for its origin.

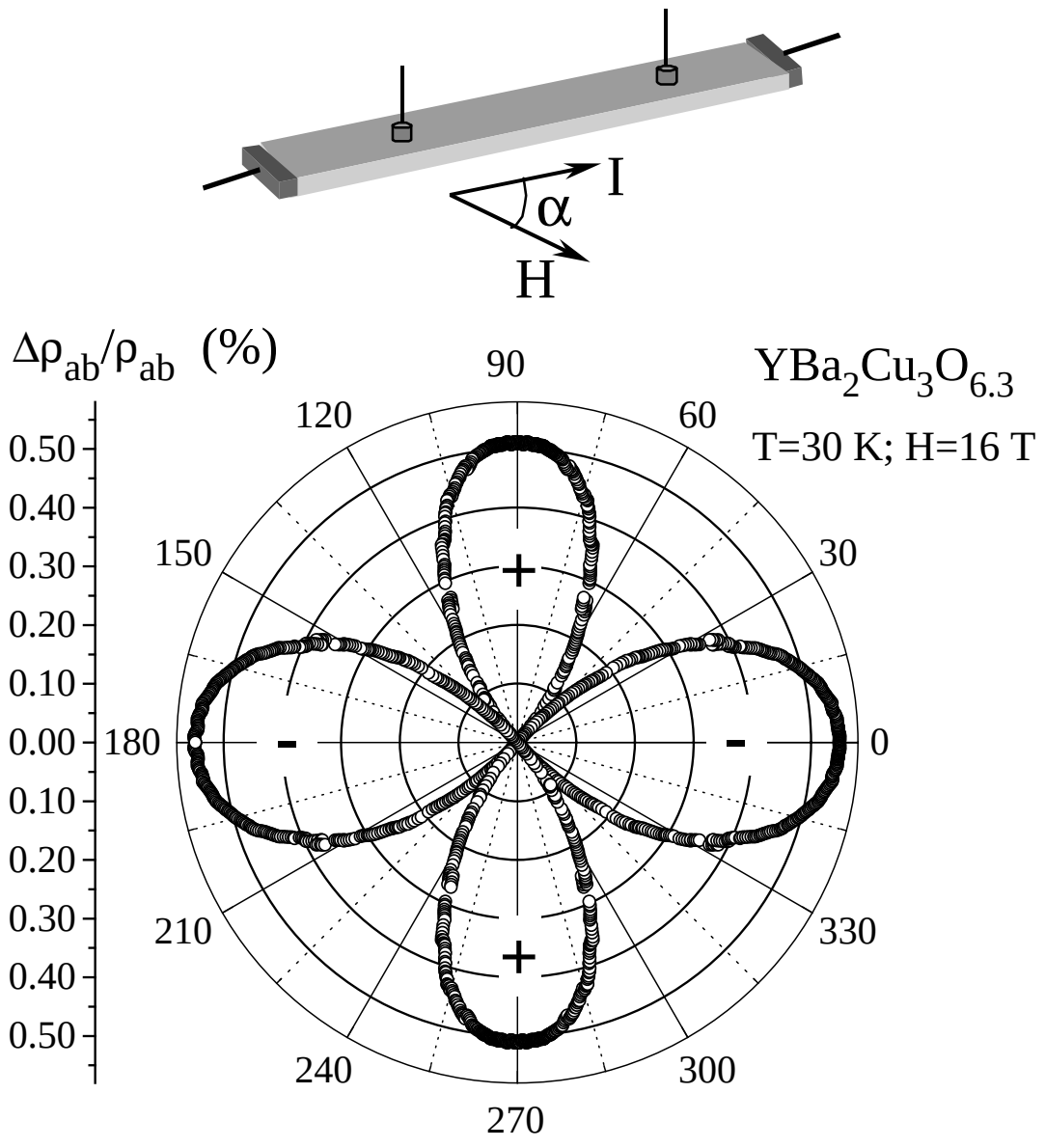

**Fig. 2.** Angular dependence of the MR (H//ab; H=16 T); the sign of MR is indicated.

When the magnetic field is turned in the plane to become perpendicular to the current [$H//ab$; $H\perp I$], the low-field MR term just switches its sign, retaining its magnitude and the threshold-field value [7]. MR measurements performed upon rotating $H$ within the $ab$ plane reveal a striking anisotropy with a "$d$-wave"-like symmetry, i.e. $\Delta\rho_{ab}/\rho_{ab}$ changes from negative at $\alpha$=0 to positive at $\alpha$=90$^o$, being zero at about 45$^o$, Fig.2. It is worth noting that the low-field MR feature is not observed at all when the magnetic field is applied along the $c$-axis.

The most intriguing peculiarity of the low-field MR appears at temperatures below ~25 K, where the $H$-dependence of $\rho_{ab}$ becomes irreversible. Figure 3 shows the low-field MR term measured for $H\perp I$ (for clarity, the background MR, $\gamma H^2$, determined at high fields is subtracted: $\Delta\rho_{ab}/\rho_{ab}=(\Delta\rho_{ab}/\rho_{ab})^*+\gamma H^2$). Initially the irreversibility appears as a small hysteresis on the MR curve, but upon cooling to 10 K it becomes much more pronounced (the MR peaks are shifted from $H$=0 and strongly suppressed). We note that the first field sweep which starts at $\Delta\rho_{ab}/\rho_{ab}$=0 differs significantly from the subsequent ones. The salient point here is that the resistivity does not return to its initial value after removing the magnetic field; hence, the system acquires a memory. The application of the magnetic field at low $T$ introduces a persistent resistivity anisotropy to the $CuO_2$ planes.

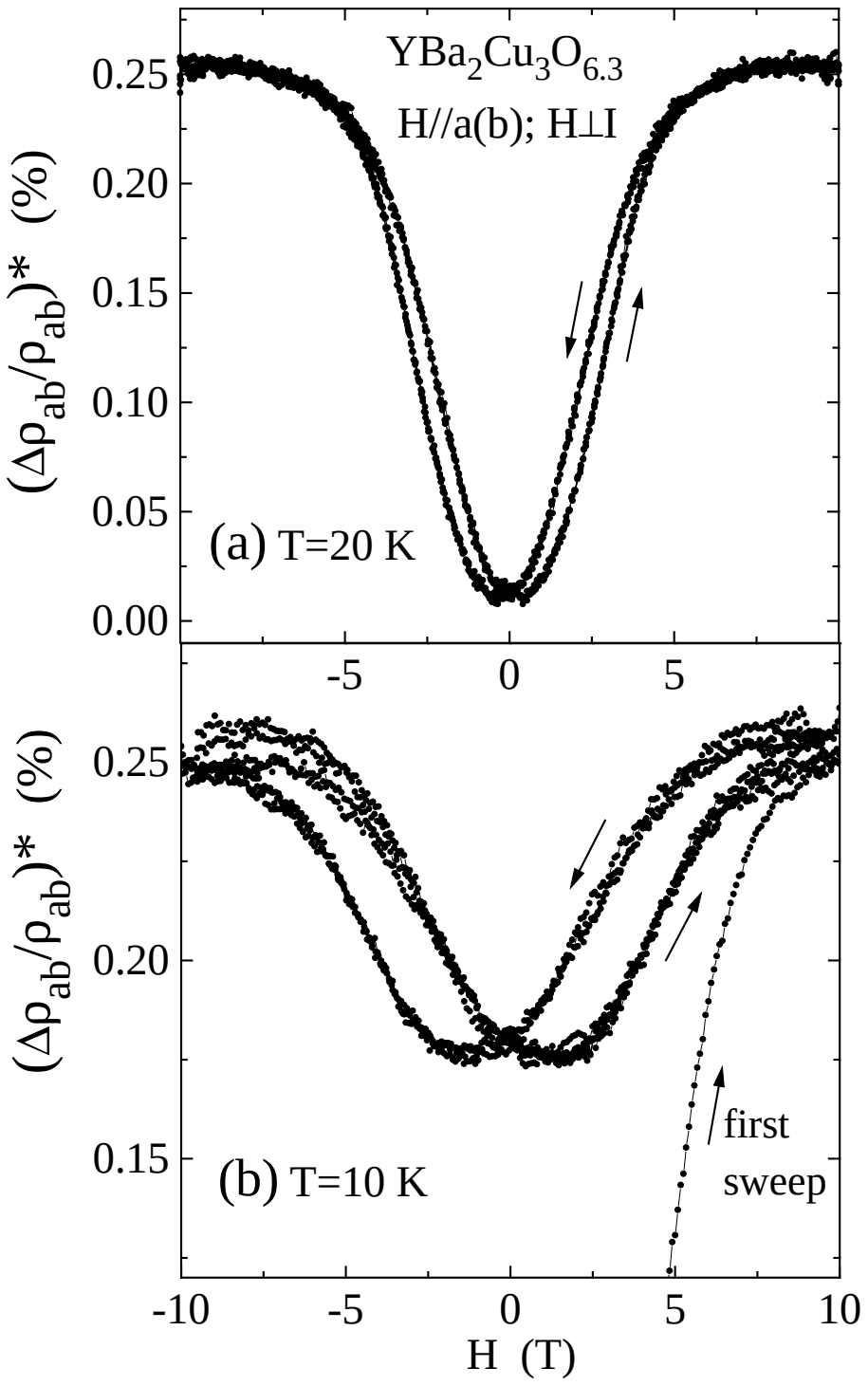

**Fig. 3.** The low-field MR component. Each curve contains data of 4 field sweeps performed at a rate of 1 T/min.

The picture would be incomplete without data on the transport between $CuO_2$ planes. It was shown that in antiferromagnetic $YBa_2Cu_3O_{6+x}$ below $T_N$, the suppression of spin fluctuations by the magnetic field results in a large positive out-of-plane MR [8]. Figure 4 shows a remarkable MR behavior produced by a superposition of the negative low-field MR feature on the positive $\gamma H^2$ background in a sample with $T_N \geq 300$ K.

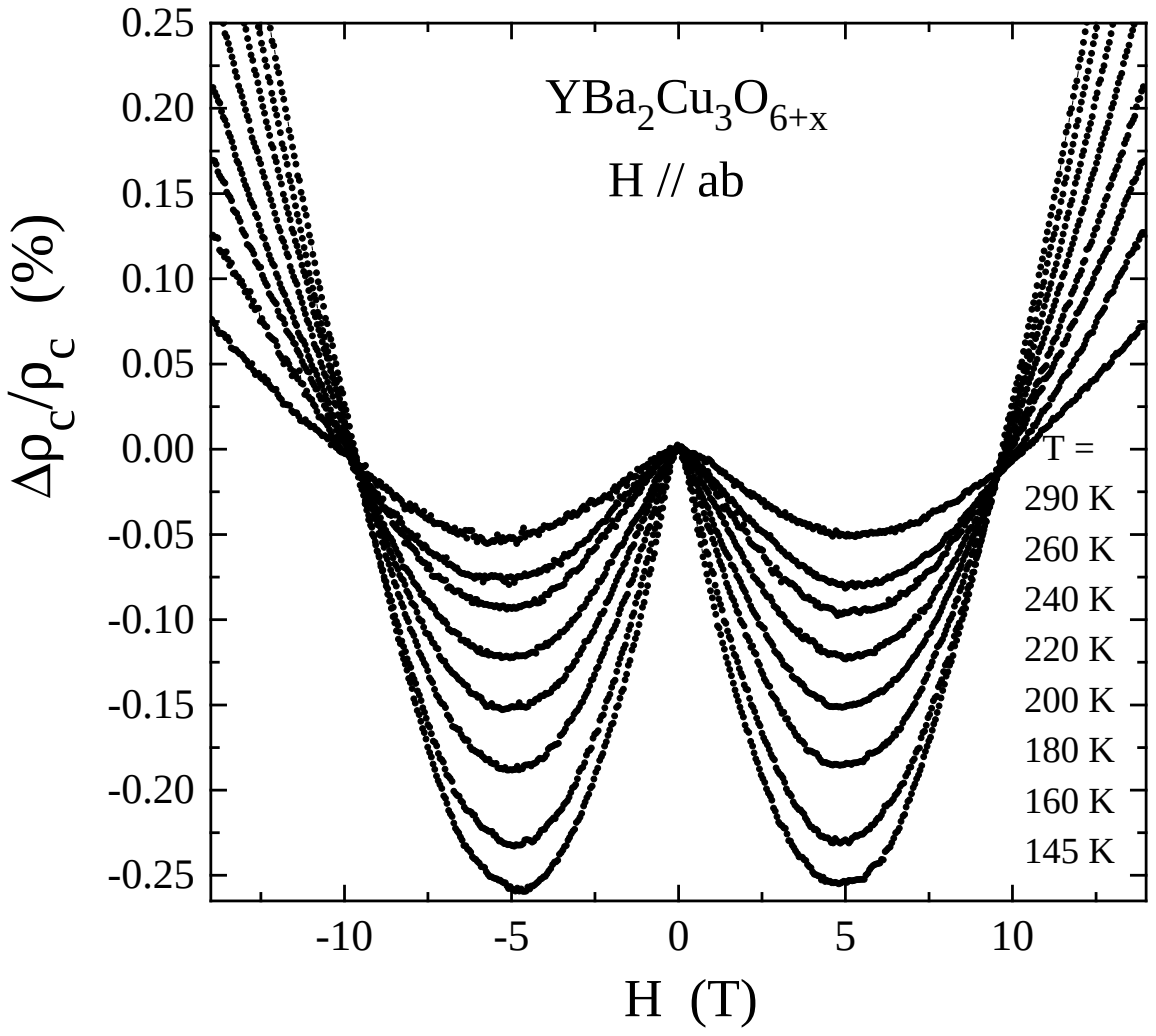

**Fig. 4.** Transverse out-of-plane MR of $YBa_2Cu_3O_{6+x}$.

It is very difficult to understand the MR anomalies presented here, especially the "memory effect", without considering an inhomogeneous state or a superstructure in the $CuO_2$ planes instead of a uniform AF state. The picture of charged "stripes" in the $CuO_2$ planes allows one to account for all the observed MR peculiarities, by assuming that the magnetic field gives rise to a directional ordering of the stripes [7]. Actually, the aligning of stripes with confined carriers moving along would change the current paths and introduce the in-plane anisotropy. The rotation of stripes by the magnetic field gives an excellent explanation for the in-plane MR with the "*d*-wave"-shaped angular dependence. Within this concept, the threshold field of several Tesla is presumably coming from the establishment of the directional order of the stripes. Though an explanation of the out-of-plane MR feature is not so straightforward, one can imagine that by adjusting the direction of stripes in neighboring $CuO_2$ planes the magnetic field increases the overlapping both between the stripes in the real space and between their quasi-1D carriers in the *k*-space, enhancing the probability of the electron hopping. As the temperature is lowered, it is expected that the stripe dynamics slows down and the magnetic domain structure in the $CuO_2$ planes is frozen, forming a cluster spin glass. The spin-glass transition temperature has been reported to be about 20-25 K for the AF compositions [9], which is in good agreement with the temperature where the hysteretic MR behavior is found.

In summary, a variety of unusual MR features found in heavily underdoped $YBa_2Cu_3O_{6+x}$ have provided new information on the conducting "stripes" in the $CuO_2$ planes. The MR behavior implies that the stripes couple to the external magnetic field and undergo topological ordering at fields of the order of a few T. Upon cooling below ~20 K the dynamics of stripes gets slower and the directional order of the stripes becomes persistent, giving rise to a "memory effect" in the resistivity. These findings show that the magnetic field can be used as a tool to manipulate the striped structure and open a possibility to clarify the electron dynamics within the stripes.

**Acknowledgments.** A.N.L. gratefully acknowledges the support from JISTEC.

---

* On leave from Institute of Inorganic Chemistry, Lavrentyeva-3, 630090 Novosibirsk, Russia.